\documentclass[journal]{IEEEtran}
\ifCLASSINFOpdf
\else
\fi
\usepackage{cite}
\usepackage{amsmath}
\usepackage{graphicx}
\usepackage{epstopdf}
\usepackage{subfigure}
\usepackage{multirow}
\usepackage{makecell}
\usepackage{etoolbox}
\usepackage{color}
\newcommand{\tabincell}[2]{\begin{tabular}{@{}#1@{}}#2\end{tabular}}

\makeatletter
\patchcmd{\@makecaption}
  {\scshape}
  {}
  {}
  {}
\makeatletter
\patchcmd{\@makecaption}
  {\\}
  {.\ }
  {}
  {}
\makeatother
\def\tablename{Table}

\begin{document}
\title{Detect and remove watermark in deep neural networks via generative adversarial networks}
\author{Haoqi Wang,
        Mingfu Xue,
        Shichang Sun,
        Yushu Zhang,
        Jian Wang,
        and~Weiqiang~Liu
\thanks{H. Wang, M. Xue, S. Sun, Y. Zhang, J. Wang are with the College of Computer Science and Technology, Nanjing University of Aeronautics and Astronautics, Nanjing, 211106, China.
W. Liu is with the College of Electronic and Information Engineering, Nanjing University of Aeronautics and Astronautics, Nanjing, 211106, China.}
\markboth{Journal of \LaTeX\ Class Files,~Vol.~14, No.~8, August~2015}}%

\maketitle

\begin{abstract}
Deep neural networks (DNN) have achieved remarkable performance in various fields. However, training a DNN model from scratch requires a lot of computing resources and training data.
It is difficult for most individual users to obtain such computing resources and training data.
Model copyright infringement is an emerging problem in recent years.
For instance, pre-trained models may be stolen or abuse by illegal users without the authorization of the model owner.
Recently, many works on protecting the intellectual property of DNN models have been proposed.
In these works, embedding watermarks into DNN based on backdoor is one of the widely used methods.
However, when the DNN model is stolen, the backdoor-based watermark may face the risk of being detected and removed by an adversary.
In this paper, we propose a scheme to detect and remove watermark in deep neural networks via generative adversarial networks (GAN). We demonstrate that the backdoor-based DNN watermarks are vulnerable to the proposed GAN-based watermark removal attack.
The proposed attack method includes two phases.
In the first phase, we use the GAN and few clean images to detect and reverse the watermark in the DNN model.
In the second phase, we fine-tune the watermarked DNN based on the reversed backdoor images.
Experimental evaluations on the MNIST and CIFAR10 datasets demonstrate that, the proposed method can effectively remove about 98\% of the watermark in DNN models, as the watermark retention rate reduces from 100\% to less than 2\% after applying the proposed attack.
In the meantime, the proposed attack hardly affect the model's performance.
The test accuracy of the watermarked DNN on the MNIST and the CIFAR10 datasets drops by less than 1\% and 3\%, respectively.
\end{abstract}
\IEEEpeerreviewmaketitle
\section{Introduction}
\IEEEPARstart{I}{n} recent years, deep neural networks (DNN) have achieved remarkable performance in many fields, such as face recognition, autonomous driving, natural language processing.
However, training a DNN model is expensive because it requires a lot of training data and computing resources.
It is extremely difficult for most individual users to train high-performance DNN models.
To this end, machine learning as a service (MLaaS) \cite{RibeiroGC15} has become an emerging business paradigm.
However, the copyright of DNN models may be infringed by malicious users.
For instance, unauthorized users may resell illegally obtained DNN models \cite{Adi2018TurningYW} or provide the pirated service based on pirated DNN models \cite{Zhang2018ProtectingIP}.
Protecting the copyright of the DNN model has great commercial value, and it has aroused widespread concerns.

A variety of methods \cite{Zhang2018ProtectingIP,Adi2018TurningYW,Merrer2019AdversarialFS,Uchida2017EmbeddingWI} have been proposed to protect the intellectual property (IP) of deep neural networks.
Among them, the backdoor-based watermarking method is one of the most popular methods.
In the backdoor based watermarking method, the model owner first injects a specific watermark trigger pattern, such as logo pattern or noise pattern \cite{Zhang2018ProtectingIP}, into clean images to create watermark trigger keys.
Then, the model owner embeds the watermark into the DNN model by using watermark trigger keys and some incorrect labels.
During copyright verification, the model owner can send watermark trigger keys to a suspicious DNN model to verify whether the model is a pirated model.

Recent works \cite{abs-1911-07205, Liu2021RemovingBW, Shafieinejad2019OnTR, Aiken2020NeuralNL, Yang2019EffectivenessOD} have shown that the backdoor-based watermarking method is vulnerable to watermark removal attacks.
Chen \textit{et al.} \cite{abs-1911-07205} proposed a fined-tuning based method, named REFET, to remove the watermark in deep neural networks. They showed that the attacker can use a well-designed learning rate schedule to fine-tune the watermarked model to remove the watermark.
Liu \textit{et al.} \cite{Liu2021RemovingBW} designed a data augmentation based method, which can remove the watermark in DNN models when a small amount of training data is available.
The data augmentation operation is used to minimize the influence of watermark triggers on the DNN model, thereby erasing the watermark in the model.
According to the attacker's knowledge of the watermarked DNN model, Shafieinejad \textit{et al.} \cite{Shafieinejad2019OnTR} proposed two attacks, a black-box attack and a white-box attack, to remove the backdoor-based watermark in the DNN.
Aiken \textit{et al.} \cite{Aiken2020NeuralNL} proposed a neural network laundering method based on Neural Cleanse \cite{Wang2019NeuralCI}.
Without knowing the size and shape of the watermark, the attacker can remove the watermark in the DNN through a three-step process, namely watermark recovery, neuron resetting, and model retraining.
Yang \textit{et al.} \cite{Yang2019EffectivenessOD} demonstrated that, the DNN's watermarking can be removed by the distillation method.
The methods \cite{abs-1911-07205, Liu2021RemovingBW} both require a lot of training data.
Thus, it is hard for these methods to be deployed in real-world scenarios.

In this paper, we propose a novel two-phase watermark removal method, which only requires few clean images.
In the first phase, we use GAN \cite{Goodfellow2014GenerativeAN} to determine whether or not there is a watermark in the DNN model.
Specifically, we first use GAN to generate perturbations to simulate the watermark trigger patterns in images.
Then, we input the simulated watermark trigger images to the watermarked DNN.
Based on the output of the watermarked DNN, GAN will modify the perturbed images gradually until finding the watermark.
Third, we use the perturbation size to determine whether the DNN model contains a watermark.
Additionally, images of different classes will all be perturbed.
If the perturbation size of a certain class is apparently less than that of other classes, then it is considered that there is a watermark in the model.
In the second phase, we fine-tune the watermarked DNN model with the reversed backdoor images to remove the watermark in the DNN model.
We have performed experiments on the MINST \cite{Deng2012TheMD} and the CIFAR10 \cite{Krizhevsky2009LearningML} datasets to evaluate the proposed method.
The experimental results show that, the test accuracy of the watermarked model trained on the CIFAR10 dataset only drops by less than 3\%, while the watermark retention rate of this model drops from 100\% to less than 2\%.
For the watermarked model trained on the MINST dataset, the watermark retention rate after the proposed is less than 0.1\%, and the test accuracy of this model only drops by less than 1\%.
Thus, the proposed method can effectively remove the watermark in the DNN model with a very small impact on the accuracy of the model.

The contributions of this paper are summarized as follows:
\begin{itemize}
\item[$\bullet$]
We proposed a novel method to remove the DNN watermark using the GAN.
In the proposed method, first, we utilize perturbations to simulate the watermark trigger pattern in watermarked images.
We use GAN to generate perturbations, such that the watermarked model can misclassify the disturbed images with the maximal probability.
Second, we utilize the reversed watermark trigger images to fine-tune the watermarked DNN to remove the watermark.
\item[$\bullet$]
The proposed GAN-based attack can detect the watermark based on the process of reversing watermark trigger patterns.
The attacker can obtain the precise target class of the backdoor-based watermarking, and the approximate position and rough shape of the watermark trigger pattern.
\item[$\bullet$]
We evaluated the proposed method on the MNIST \cite{Deng2012TheMD} and CIFAR10 \cite{Krizhevsky2009LearningML} datasets.
The experimental results show that the proposed method can effectively remove the watermark in the DNN model, while having a limited impact on the performance of the DNN model.
\end{itemize}

The organization of this paper is as follows.
Section \ref{relatedWork} introduces the related work, including the DNN watermarking works and the existing watermark removal works.
Section \ref{proposedMethod} elaborates on the proposed GAN-based watermark removal attack from two aspects: watermark reversing and watermark removal.
Section \ref{experiment} evaluates the proposed attack.
Section \ref{conclusion} summarizes this paper.

\section{Related work} \label{relatedWork}
In this section, we review the DNN watermarking methods and watermark removal attacks.

\subsection{DNN watermarking method}
In the field of DNN copyright protection, many works have been proposed \cite{Uchida2017EmbeddingWI, Adi2018TurningYW, Zhang2018ProtectingIP, Merrer2019AdversarialFS}.
Uchida \textit{et al.} \cite{Uchida2017EmbeddingWI} proposed to use a binary string as the watermark and embed it to the selected layer of the host DNN model through a parameter regularizer.
Adi \textit{et al.} \cite{Adi2018TurningYW} proposed the backdoor-based DNN watermarking method.
In this method, the abstract images and randomly selected classes are treated as the trigger set, and the watermark is embedded into the DNN model by utilizing the trigger set during training.
Based on the backdoor, Zhang \textit{et al.} \cite{Zhang2018ProtectingIP} proposed three watermark generation algorithms, which are the content-based algorithm, the unrelated-based algorithm, and the noise-based algorithm.
The content-based algorithm treats the meaningful content such as logo pattern as a watermark, and embeds the watermark into the DNN model through the backdoor technique.
Similarly, the unrelated-based algorithm treats irrelevant samples or images as a watermark, and the noise-based algorithm treats the noise as a watermark.
Merrer \textit{et al.} \cite{Merrer2019AdversarialFS} leveraged a set of adversarial samples, named the watermark key set, to query and verify whether a remote model has been watermarked.

\subsection{Watermark removal method}
At present, five works have explored feasible methods to attack the existing watermarking methods, especially the backdoor-based watermarking methods.
Chen \textit{et al.} \cite{abs-1911-07205} proposed a fine-tuning based method, called REFIT, to remove watermarks in DNN models.
REIFT utilizes unlabeled data and a learning rate schedule to fine-tune the model so as to remove the watermark.
Liu \textit{et al.} \cite{Liu2021RemovingBW} proposed a watermark removal attack, named WILD, which is effective to remove the watermark when the amount of training data is limited.
In WILD, data augmentation and a distribution alignment method are used together to minimize the trigger effect of the watermark triggers.
Shafieinejad \textit{et al.} \cite{Shafieinejad2019OnTR} proposed two attacks, a black-box attack and a white-box attack, to remove the backdoor-based watermarking, when the attacker has different knowledge of the watermarked DNN.
The black-box attack is based on the model extraction attack \cite{TramerZJRR16}, while the white-box attack is based on regularization and fine-tuning.
Aiken \textit{et al.} \cite{Aiken2020NeuralNL} proposed a neural network laundering method.
Different from existing watermark removal methods, they first performed reverse engineering to reconstruct the watermarked images, then combined neuron pruning and model retraining to remove watermarks in watermarked DNN models.
Yang \textit{et al.} \cite{Yang2019EffectivenessOD} proposed to utilize a distillation method to remove the watermark.
Since the watermark in DNN is independent to DNN's main learning task, it will be vulnerable to distillation and will not be memorized by the distilled model.

The above watermark removal methods \cite{abs-1911-07205, Liu2021RemovingBW} are hard to be deployed in real-world scenarios.
The watermark removal method proposed in \cite{abs-1911-07205} relies on a carefully designed learning rate schedule, but designing an effective learning rate plan requires a lot of trial and error.
The method proposed in \cite{Liu2021RemovingBW} uses data augmentation technology to reduce the amount of required training data, but the amount of data is still very large for an attacker.

The difference between the work in this paper and the existing watermark removal methods \cite{abs-1911-07205, Liu2021RemovingBW} are as follows:
\begin{itemize}
\item[1)]
We use GAN to determine whether there is a watermark in the model.
If there is no watermark in the DNN model, the attacker will not alter the model any more.
\item[2)]
Compared with WILD \cite{Liu2021RemovingBW} and REFIT \cite{abs-1911-07205}, our method requires less training data, which ensures that the proposed method is more easy to be deployed in the real-world scenarios.
\end{itemize}

\section{The proposed method}\label{proposedMethod}
\subsection{Overview}
\begin{figure*}[htbp]
\centering
\includegraphics[scale=0.7]{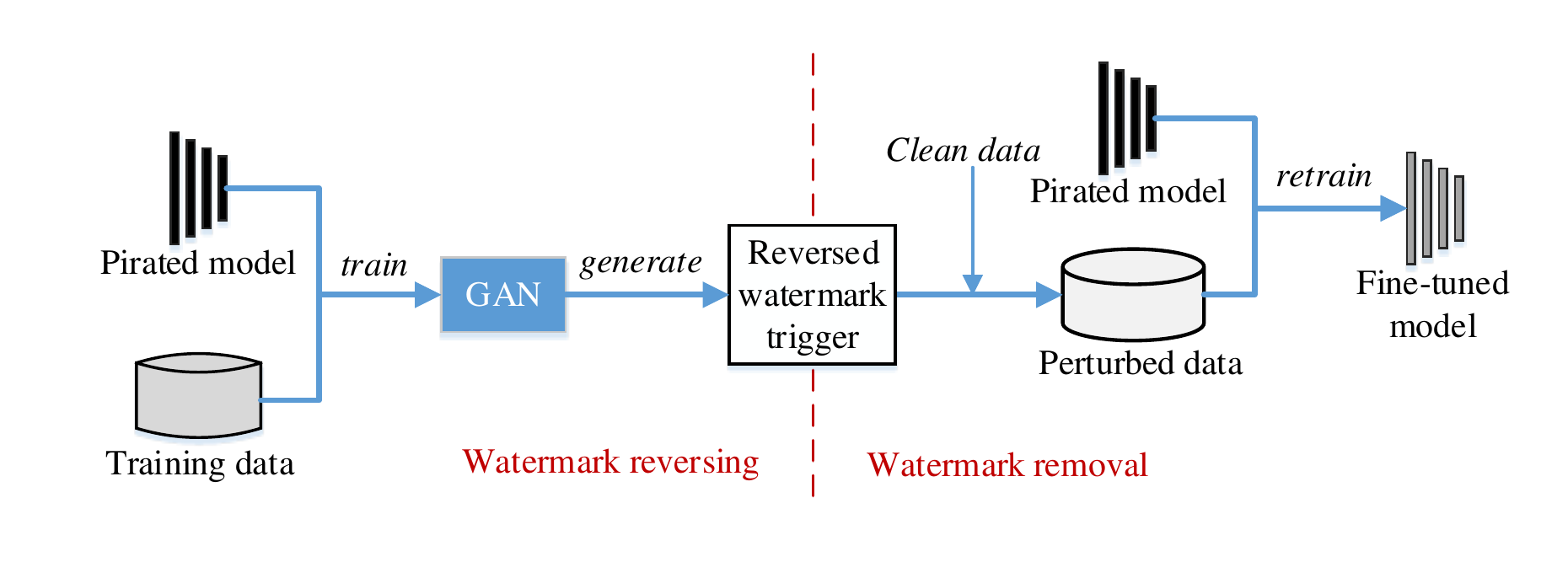}
\caption{Overview of the proposed watermark removal attack.}
\label{fig1}
\end{figure*}

The overview of the proposed attack is shown in Fig. \ref{fig1}. It can be divided into two phases, i.e., watermark reversing and watermark removal.
\begin{itemize}
\item[1)]\textbf{Watermark reversing}:
We attempt to mimic the watermark trigger pattern with perturbations generated by GAN.
Given a small amount of training data and a pirated DNN model, we train a GAN to output special perturbation patterns to reverse the watermark.
During the training process, we first assume that there is a watermark in the pirated DNN model, and then enumerate every class to find the watermarked class.
Third, a metric named perturbation size is used to determine which class is watermarked.
The perturbation size indicates the $L_2$ distance between the perturbed image and the original image.
If the perturbation size of a certain class is apparently less than the perturbation size of other classes, then we consider that such unique class is watermarked.
Lastly, the perturbations generated by GAN are considered as the reversed watermark trigger pattern.
\item[2)]\textbf{Watermark removal}:
We adopt a fine-tuning method to remove the watermark in the DNN. Unlike the existing fine-tuning methods, we only use reversed watermark images to fine-tune the watermarked model such that the model forgets the learned watermark.
\end{itemize}

\subsection{Watermark reversing}
\begin{figure}[htbp]
\centering
\includegraphics[scale=0.52]{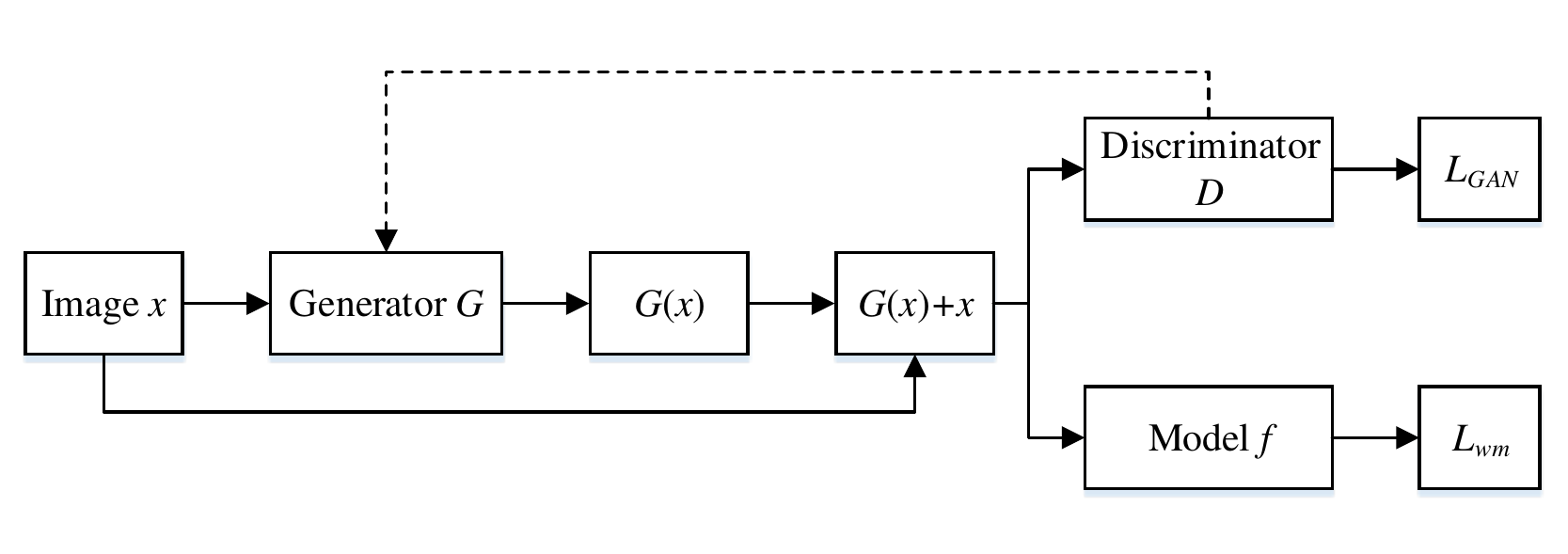}
\caption{The structure and workflow of GAN in the proposed attack, where $G$ denotes the generator and $G$ denotes the discriminator.}
\label{fig2}
\end{figure}

In the watermark reversing phase, we adopt the AdvGAN \cite{XiaoLZHLS18} to infer whether there is a watermark in the pirated DNN model.
The AdvGAN \cite{XiaoLZHLS18} consists of a generator $G$, a discriminator $D$, and the target DNN model $f$, which is shown in Fig. \ref{fig2}.
First, a clean image is fed into the generator $G$, which will output a specific perturbation $G(x)$.
Then, the perturbation $G(x)$ is added to the clean image to craft an adversarial example $G(x)+x$.
Third,  $G(x)+x$ will be input to the discriminator and the target DNN model simultaneously.
For the discriminator $D$, it is used to distinguish sample $G(x)+x$ from the real sample $x$.
The output of $D$ will be fed back to the generator $G$ to encourage $G$ to generate a more indistinguishable sample in the next iteration.
To generate a more realistic adversarial example,  we optimize both the generator and the discriminator by the loss $L_{GAN}$ \cite{XiaoLZHLS18}:
\begin{equation}
\label{equ1}
{L_{GAN}} = MSE(D(x),1) + MSE(D(G(x) + x),0)
\end{equation}
where $MSE$ represents the mean square error, and $G(x)+x$ represents the perturbed sample. $L_{GAN}$ encourages the generated adversarial example to be close to the original image.

For the target DNN model $f$, it takes the sample $G(x)+x$ as input and outputs the loss $L_{wm}$. In the proposed attack, $L_{wm}$ is calculated by \cite{Zhu2020GangSweepSO}:
\begin{equation}
{L_{{\rm{wm}}}} = \max (\max \{ f{(G(x) + x)_i}:i \ne t\}  - f{(G(x) + x)_t},0)
\end{equation}
where $t$ is the true class of the image $x$, and $i$ represents other classes except for $t$.
$L_{wm}$ encourages the generated adversarial sample $G(x)+x$ to be classified into an incorrect class $i$.
In addition, by optimizing the distance between the ground truth label and another label with the highest probability, the generated perturbation can shift in the direction that is most likely to be the backdoored class.

In order to constrain the magnitude of the generated perturbation, we use the $L_2$ distance as follows:
\begin{equation}
\label{equ3}
{L_{pert}} = ||G(x)|{|_2}
\end{equation}
Then, the overall objective function for attacking the target DNN model is as follows:
\begin{equation}
\label{equ4}
L = {\lambda _1}{L_{wm}} + {\lambda _2}{L_{pert}}
\end{equation}
where $\lambda _1$ and $\lambda _2$ are two hyperparameters that are used to balance each objective item.

Note that, in the proposed attack, we call $L_{pert}$ the perturbation size and use it to determine whether there is a watermark in the target DNN.
Specifically, given a DNN model $f$ and a small number of validation images, we first assume an arbitrary class $c$ is the target class of the watermarked DNN. Then, we calculate the expected value of $L_{pert}$ on the target class $c$ with the validation images.
By enumerating all classes, if the value of perturbation size on a certain class $c$ is significantly smaller than that on other classes, we consider that $c$ is the true target class $b$ of the backdoor-based watermarking.
We refer to perturbation size on the true target class $b$ as the perturbation outlier, and the value of perturbation size on any other class as the normal perturbation size.
In general, the value of perturbation outlier is less than the pre-defined threshold $T$, while the value of normal perturbation size is greater than $T$.
Empirically, the value of $T$ is in the range 9$\sim$10, which derives from the experimental results in Section \ref{parameterDiscuss}.
In addition to the target class $b$, we can also obtain the perturbation from the generator $G$. The generated perturbation $G(x)$ is considered as the reversed watermark trigger.

\textbf{Analysis on GAN-based perturbation generation}.
In the proposed attack, we attempt to simulate the watermark trigger in the watermarked DNN through perturbations.
The generative adversarial networks is able to generate perturbations without greatly damaging the image, thus we leverage the GAN architecture to generate perturbations and reverse watermark triggers.
The target DNN model $f$ is also used to assist the generation of perturbations.
Based on the backpropagation process of the loss $L_{wm}$ from model $f$, the generator $G$ tends to craft perturbations in the opposite direction of the ground truth label.
These generated perturbations could be randomly distributed or clustered together.
From our perspective, if there is no watermark in the model $f$, the perturbation generated by GAN is randomly distributed.
However,  if there exists a watermark, the perturbation generated by GAN can be more inclined to trigger the watermark, that is, GAN will generate a rough perturbation pattern at the position of the watermark trigger in clean images.

\begin{figure*}[htbp]
\centering
\subfigure[]
{
    \begin{minipage}[b]{.15\linewidth}
        \centering
        \includegraphics[height=2.5cm, width=2.5cm]{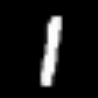} \\
        \hspace{0.15cm}
        \includegraphics[height=2.5cm, width=2.5cm]{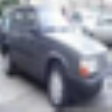}
    \end{minipage}
}
\subfigure[]
{
    \begin{minipage}[b]{.15\linewidth}
        \centering
        \includegraphics[height=2.5cm, width=2.5cm]{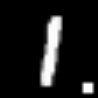} \\
        \hspace{0.15cm}
        \includegraphics[height=2.5cm, width=2.5cm]{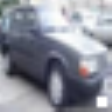}
    \end{minipage}
}
\subfigure[]
{
    \begin{minipage}[b]{.15\linewidth}
        \centering
        \includegraphics[height=2.5cm, width=2.5cm]{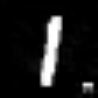} \\
        \hspace{0.15cm}
        \includegraphics[height=2.5cm, width=2.5cm]{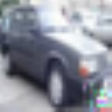}
    \end{minipage}
}
\subfigure[]
{
    \begin{minipage}[b]{.15\linewidth}
        \centering
        \includegraphics[height=2.5cm, width=2.5cm]{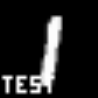} \\
        \hspace{0.15cm}
        \includegraphics[height=2.5cm, width=2.5cm]{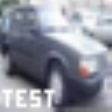}
    \end{minipage}
}
\subfigure[]
{
    \begin{minipage}[b]{.15\linewidth}
        \centering
        \includegraphics[height=2.5cm, width=2.5cm]{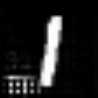} \\
        \hspace{0.15cm}
        \includegraphics[height=2.5cm, width=2.5cm]{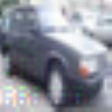}
    \end{minipage}
}
\caption{Example images on the MNIST and CIFAR10 datasets. (a) Clean images; (b) watermarked images with a white square; (c) reversed watermark images (white square); (d) watermarked images with a TEST pattern; (e) reversed watermark images (TEST pattern).}
\label{fig3}
\end{figure*}

Fig. \ref{fig3} shows several example images, including clean images, watermarked images, and reversed watermark images. As shown in Fig \ref{fig3}(b) and Fig. \ref{fig3}(c), on the MNIST \cite{Deng2012TheMD} dataset, the white square trigger reversed by GAN is very similar to the real white square trigger. On the CIFAR10 \cite{Krizhevsky2009LearningML} dataset, the reversed watermark trigger does not resemble the real watermark trigger, but the position of the watermark trigger is roughly correct in the reversed image.
As shown in Fig \ref{fig3}(d) and Fig. \ref{fig3}(e), the reverse watermark trigger is not very similar to the real watermark trigger (TEST pattern), but in the reverse image, the position of the reverse watermark pattern is still close to the position of the real trigger.

\subsection{Watermark removal}
The effect of watermark removal relies on how similar the reversed watermark pattern is to the real watermark trigger.
To this end, we optimize the previously reversed watermark pattern by selecting the pattern with minimal perturbation size.
In order to remove the watermark in the DNN model, a fine-tuning based pipeline is performed by the following steps.
\begin{enumerate}
  \item Superimpose the selected watermark pattern (i.e., the reversed watermark trigger) on clean images. These combined images are used as training samples.
  \item Assign the correct labels (i.e., the ground truth labels) to the training samples.
  \item fine-tune or retrain the watermarked DNN model with these training samples.
\end{enumerate}

This fine-tuning process, also called the unlearning \cite{Wang2019NeuralCI}, can effectively remove the watermark by using a small amount of training data.
In the proposed method, at most 10\% training data of the whole dataset is required.
In addition, our watermark removal attack will not degrade the performance of the watermarked DNN, as the fine-tuning will automatically compensate for the accuracy loss of the watermarked DNN.

\section{Experiment } \label{experiment}
In this section, we will evaluate the proposed watermark removal attack. First, we introduce the experimental setup, including the dataset, deep neural networks, watermark triggers, and evaluation metrics.
Then, we perform experimental evaluation and analysis on the performance of the proposed method.
Finally, we compare the proposed attack with two existing watermark removal works \cite{Liu2021RemovingBW, abs-1911-07205}.

\subsection{Experimental setup}
\textbf{Datasets}.
We use the MNIST \cite{Deng2012TheMD} and CIFAR10 \cite{Krizhevsky2009LearningML} datasets to evaluate the performance of the proposed method. MNIST is a handwritten digital image dataset consisting of grayscale images with 10 classes \cite{Deng2012TheMD}.
The size of each image in the MNIST dataset is $28 \times 28$.
CIFAR10 is an object classification dataset, which also contains 10 classes \cite{Krizhevsky2009LearningML}.
Each image is a color image with a size of $32 \times 32$.

\textbf{Deep neural networks}. We adopt the LeNet-5 \cite{LeCun1998GradientbasedLA} and ReNet-18 \cite{He2016DeepRL} architectures for experimental evaluations.
The LeNet-5 model is trained on the MNIST dataset for 80 epochs to embed the watermark, and the ResNet-18 model is trained on the CIFAR10 dataset for 100 epochs to embed the watermark.

\textbf{Watermark triggers}.
We adopt two widely used patterns, the white square and the ``TEST'' logo as watermark triggers. Both of the two patterns are added in the lower right corner of each image. We sample 5\% of the data from the training set of the MNIST and CIFAR10 datasets as the watermark trigger samples.
The selected two types of watermarks are embedded into the DNN models through these watermark trigger samples and the assigned target label ``7'' during training.

\textbf{Evaluation metrics}.
We adopt the following three metrics to evaluate the effectiveness of our method on removing DNN watermarks.
\begin{itemize}
\item[1)]\textbf{Test accuracy}: We evaluate the performance of DNN models by calculating the test accuracy before and after the proposed  attack.
\item[2)]\textbf{Watermark retention rate}\cite{Liu2021RemovingBW}:
We use the watermark retention rate to measure the watermark removal effect after applying the proposed attack.
The watermark retention rate represents the probability of correctly classified watermark samples in all watermark samples.
Assuming that the number of correctly classified watermark samples is $S_y$, and the total number of watermark samples is $S$, then the watermark retention rate is calculated by ${S_y}/S$.
\item[3)] \textbf{Perturbation size}:
We use perturbation size $L_{pert}$ to determine whether or not a DNN model has a watermark on a certain class.
If the value of $L_{pert}$ on a certain class is less than the threshold $T$, then there is a watermark in the DNN model.
The perturbation size $L_{pert}$ is calculated by Formula \eqref{equ3}.

\end{itemize}
\subsection{Effectiveness of watermark removal}
In this section, we evaluate the proposed attack on the MNIST \cite{Deng2012TheMD} and CIFAR10 \cite{Krizhevsky2009LearningML} datasets respectively.
Two types of watermarks (i.e., white square and ``TEST'' pattern) are embedded in LeNet-5 \cite{LeCun1998GradientbasedLA} and ReNet-18 \cite{He2016DeepRL} models, respectively.
The attacker applies the proposed attack to detect and remove the watermark in the DNN model.

The effectiveness of the proposed attack is shown in Table \ref{tab1}.
Here, the basic test accuracy and basic watermark retention rate indicate the performance of the DNN model before the proposed attack.
It can be seen that, on the MNIST dataset, the watermark retention rate of the watermarked LeNet-5 drops from over 99\% to about 1\%, while the test accuracy is only reduced by less than 1\%. Similarly, on the CIFAR10 dataset, the watermark retention rate of ResNet-18 drops from over 99\% to about 1.4\%, while the test accuracy only drops by less than 3\%.
In addition, the proposed attack performs well on removing the white square watermark and the ``TEST'' pattern watermark, as the watermark removal rates of the two watermarks are both greater than 98\%.
Therefore, our proposed attack can effectively remove watermarks in DNN models without affecting the test accuracy of the DNN.

The reason why the proposed method is effective in removing watermarks in DNN models is summarized as follows.
To remove the watermark, we first perform a GAN-based watermark reversing process, where the watermarked DNN mistakes the perturbation generated by GAN as a watermark trigger pattern.
In the model fine-tuning stage, when the reversed watermark trigger images are remarked as the correct class labels during training, the watermarked model tends to forget the previously learned watermark. Thus, the watermark will be successfully removed after applying the proposed attack. In addition, the fine-tuning operation enables the DNN model to maintain its test accuracy.

\begin{table*}[htbp]
  \renewcommand\arraystretch{1.3}
  \centering
  \caption{Test accuracy and watermark retention rate of the DNN model before and after the watermark removal attack}
    \begin{tabular}{|c|c|c|c|c|c|}
    \hline
    \textbf{Dataset} & \textbf{Watermark types} & \textbf{\tabincell{c}{Basic \\ test accuracy}}  & \textbf{\tabincell{c}{Basic watermark \\retention rate}} & \textbf{Test accuracy} & \textbf{\tabincell{c}{Watermark \\retention rate}} \\
    \hline
    \textbf{\multirow{2}[3]{*}{MNIST \cite{Deng2012TheMD}}} & White square & 99.59\% & 99.93\% & 98.67\% & 1\% \\
    \cline{2-6}      & TEST pattern & 99.34\% & 99.99\% & 98.57\% & 1.2\% \\
    \hline
    \textbf{\multirow{2}[3]{*}{CIFAR10 \cite{Krizhevsky2009LearningML}} } & White square & 86.53\% & 99.96\% & 84.08\% & 1.42\% \\
    \cline{2-6}      & TEST pattern & 86.11\% & 99.99\% & 83.44\% & 1.4\% \\
    \hline
    \end{tabular}
  \label{tab1}
\end{table*}

\subsection{Parameter discussion}\label{parameterDiscuss}
In this section, we discuss the influence of different parameters on the performance of the proposed attack.
In the first stage of the proposed attack, the robustness of our method to different target classes is discussed.
In the second stage of the proposed attack, the impact of different training data and epochs on the proposed attack is discussed.

We first discuss the impact of different target classes on the defined perturbation outliers.
To this end, we perform four groups of experiments on the MNIST dataset, and the selected target classes for embedding watermarks are 1, 4, 7, and 9, respectively.
The perturbation size and perturbation outliers on different classes are shown in Table \ref{tab2}.
It can be seen that, in each group of experiments, the perturbation outlier (in red color) only appears on the target class.
For instance, in the first group of experiments (shown in the second row of Table \ref{tab2}), the perturbation size on the target class 1 (i.e., the perturbation outlier) is 7.565, while the perturbation size on other classes is in the range of 14$\sim$17.
In all four groups of experiments, the perturbation outliers are in the range of 7$\sim$8, but the normal values of perturbation sizes are in the range of 14$\sim$17.
Attackers can determine whether a DNN model has a watermark from the perturbation size.
Similarly, on the CIFAR10 dataset, we also perform four groups of experiments, and the target classes are 0, 2, 5, and 7, respectively.
The experimental results are shown in Table \ref{tab3}.
It is shown that, after the proposed attack, the perturbation size on the target class is significantly smaller than the perturbation sizes on other normal classes.
In the proposed method, the selection of the target class in the watermarked model has almost no influence on the proposed attack.
In other words, the proposed attack is robust to backdoor-based watermarking methods whose target class is randomly selected.
In addition, the threshold for distinguishing perturbation outliers from normal perturbation sizes can be selected in the range $9\sim10$.
\begin{table*}
\begin{center}
    \renewcommand\arraystretch{1.3}
    \caption{Perturbation size and perturbation outliers on different classes. Four groups of experiments are performed on the MNIST dataset, and the target classes used are 1, 4, 7, and 9, respectively. The data marked in red color represent the perturbation outliers.}
    \begin{tabular}{|c|c|c|c|c|c|c|c|c|c|c|}
    \hline
      \textbf{Class} & 0 & 1 & 2 & 3 & 4 & 5 & 6 & 7 & 8 & 9\\
      \hline
      \multirow{4}*{\textbf{\tabincell{c}{Perturbation \\ size}}}
      & 14.865 & \textcolor{red}{7.565} & 16.214 & 15.012 & 16.11 & 15.421 & 15.742 & 14.365 & 15.458 & 15.897\\
      \cline{2-11}
      & 15.14 & 14.623 & 15.562 & 15.354 & \textcolor{red}{7.569} & 14.424 & 15.286 & 16.656 & 14.195 & 16.433\\
      \cline{2-11}
      & 15.403 & 16.436 & 15.487 & 15.15 & 16.175 & 14.975 & 15.468 & \textcolor{red}{7.077} & 15.743 & 16.124\\
      \cline{2-11}
      & 15.101 & 16.109 & 15.981 & 15.008 & 15.754 & 15.343 & 15.788 & 14.412 & 15.98 & \textcolor{red}{7.961}\\
      \hline
    \end{tabular}
    \label{tab2}
\end{center}
\end{table*}

\begin{table*}
\begin{center}
    \renewcommand\arraystretch{1.3}
    \caption{Perturbation size and perturbation outliers on different classes. Four groups of experiments are performed on the CIFAR10 dataset, and the target classes used are 0, 2, 5, and 7, respectively. The data marked in red color represent the perturbation outliers.}
    \begin{tabular}{|c|c|c|c|c|c|c|c|c|c|c|}
    \hline
      \textbf{Class} & 0 & 1 & 2 & 3 & 4 & 5 & 6 & 7 & 8 & 9\\
      \hline
      \multirow{4}*{\textbf{\tabincell{c}{Perturbation \\ size}}}
      & \textcolor{red}{7.112} & 14.589 & 13.124 & 14.124 & 12.441 & 13.745 & 12.118 & 13.547 & 12.778 & 12.745\\
      \cline{2-11}
      & 12.578 & 12.734 & \textcolor{red}{7.11} & 13.254 & 12.52 & 14.112 & 12.584 & 14.257 & 13.122 & 11.245\\
      \cline{2-11}
       & 13.589 & 12.968 & 12.475 & 13.785 & 12.714 & \textcolor{red}{7.569} & 12.325 & 12.956 & 13.678 & 11.989\\
       \cline{2-11}
       & 12.449 & 12.495 & 11.973 & 13.252 & 11.955 & 15.456 & 11.545 & \textcolor{red}{7.692} & 12.214 & 12.488\\
      \hline
    \end{tabular}
    \label{tab3}
\end{center}
\end{table*}

In the second stage of the proposed attack, we discuss the impact of different number of training data and different number of epochs on the effect of watermark removal.
We set the training data used for fine-tuning to be 10\%, 5\%, and 2\% of the training set, respectively.
On the MNIST dataset \cite{Deng2012TheMD}, 6,000, 3,000, 1,200 images are used to fine-tune the watermarked DNN respectively. On the CIFAR10 dataset \cite{Krizhevsky2009LearningML}, 5,000, 2,500, 1,000 images are used to fine-tune the watermarked DNN respectively.
Moreover, we use 10 epochs, 40 epochs, and 80 epochs to fine-tune the watermark DNN to remove the watermark.
The experimental results are shown in Table \ref{tab4} and Table \ref{tab5}.
As shown in Table \ref{tab4}, when 10\% training data is used in the proposed attack, the watermark retention rate degrades from almost 100\% to below 3\%, while the test accuracy drops by 3.02\% at most.
When 5\% of the training data is used, the watermark retention rate is in the range of 2\%$\sim$4\%, and the test accuracy is reduced by up to 6\%.
When 2\% of the training data is used, the watermark retention rate is in the range of 4\%$\sim$11\%.
Thus, when about 5\% training data is used, the proposed attack has an excellent watermark removal effect, while having a small impact on the test accuracy of the DNN model.
In addition, as the training epoch increases, the watermark removal effect of the proposed attack gets better, as well as the test accuracy.
In Table \ref{tab5}, we can draw similar conclusions on the CIFAR10 dataset.
In summary, the proposed method will be affected by different number of training data and different number of epochs.

\begin{table*}
\renewcommand\arraystretch{1.3}
\begin{center}
    \caption{Test accuracy and watermark retention rate after watermark removal on the MNIST \cite{Deng2012TheMD} dataset.}
    \begin{tabular}{|c|c|c|c|c|c|c|c|}
    \hline
      \multirow{2}*{\textbf{Watermark}} & \multirow{2}*{\textbf{\makecell{Percentage of \\training data}}} & \multicolumn{3}{c|}{\textbf{\makecell{Test accuracy after watermark \\removal}}} & \multicolumn{3}{c|}{\textbf{\makecell{Watermark retention rate after\\ watermark removal}}} \\
      \cline{3-8}
      & &10 epochs	&40 epochs&80 epochs&10 epochs&40 epochs&80 epochs\\
      \hline
      \multirow{3}*{\textbf{\makecell{White\\ square}}}&10\%  & 96.79\% & 98.07\% & 98.67\% & 1.65\% & 1.57\% & 1\%\\
      \cline{2-8}
      &5\%  & 94.33\% & 97.7\% & 98.37\% & 3.2\% & 2.13\% & 1.8\%\\
      \cline{2-8}
      &2\%  & 87.92\% & 95.69\% & 96.63\% & 10.1\% & 8.42\% & 4.3\%\\
      \hline
      \multirow{3}*{\textbf{\makecell{TEST\\ pattern}}}& 10\%  & 96.32\% & 98.27\% & 98.57\% & 2.14\% & 1.99\% & 1.2\%\\
      \cline{2-8}
      &5\%  & 94.13\% & 97.33\% & 98.04\% & 3.12\% & 2.33\% & 1.2\%\\
      \cline{2-8}
      &2\%  & 87.71\% & 95.1\% & 96.68\% & 1.12\% & 8.79\% & 4.3\%\\
      \hline
    \end{tabular}
    \label{tab4}
\end{center}
\end{table*}

\begin{table*}
\begin{center}
\renewcommand\arraystretch{1.3}
    \caption{Test accuracy and watermark retention rate after watermark removal on the CIFAR10 \cite{Krizhevsky2009LearningML} dataset.}
    \begin{tabular}{|c|c|c|c|c|c|c|c|}
    \hline
      \multirow{2}*{\textbf{Watermark}} & \multirow{2}*{\textbf{\makecell{Percentage of \\training data}}} & \multicolumn{3}{c|}{\textbf{\makecell{Test accuracy after \\watermark removal}}} & \multicolumn{3}{c|}{\textbf{\makecell{Watermark retention rate after\\ watermark removal}}} \\
      \cline{3-8}
      & &10 epochs	&40 epochs&80 epochs&10 epochs&40 epochs&80 epochs\\
      \hline
      \multirow{3}*{\textbf{\makecell{White\\ square}}}&10\%  & 73.13\% & 82.18\% & 84.08\% & 12.33\% & 5.48\% & 1.42\%\\
      \cline{2-8}
      &5\%  & 53.31\% & 77.1\% & 81.6\% & 10.1\% & 5.52\% & 1.5\%\\
      \cline{2-8}
      &2\%  & 38.93\% & 62.17\% & 65.93\% & 11.03\% & 10.22\% & 8.86\%\\
      \hline
      \multirow{3}*{\textbf{\makecell{TEST\\ pattern}}}&10\%  & 74.67\% & 83.6\% & 83.44\% & 12.17\% & 6.12\% & 1.4\%\\
      \cline{2-8}
      &5\%  & 51.97\% & 75.32\% & 81.08\% & 10.4\% & 5.81\% & 1.2\%\\
      \cline{2-8}
      &2\%  & 42.93\% & 61.86\% & 67.95\% & 11.42\% & 10.86\% & 8.32\%\\
      \hline
    \end{tabular}
    \label{tab5}
\end{center}
\end{table*}

\subsection{Comparison with existing works}
We compare the proposed watermark removal method with two existing watermark removal methods, REFIT \cite{abs-1911-07205} and WILD \cite{Liu2021RemovingBW}.
Since the two works \cite{abs-1911-07205, Liu2021RemovingBW} do not provide the source codes, it is difficult for us to reproduce these two works in the experiment.
Therefore, we only use the data in the paper WILD \cite{Liu2021RemovingBW} for comparison.
10\% data of the training set is used to remove the watermark.
Table \ref{tab6} presents the compared results of the proposed watermark removal method with REFIT and WILD.

\begin{table*}[htbp]
\renewcommand\arraystretch{1.3}
\begin{center}
    \caption{Comparison of the proposed method with existing work on watermark removal. Among them, the data of REFIT \cite{abs-1911-07205} and WILD \cite{Liu2021RemovingBW} are selected from \cite{Liu2021RemovingBW}, and the watermark is based on TEST pattern.}
    \begin{tabular}{|c|c|c|c|c|c|c|}
    \hline
      \textbf{Dataset}& \textbf{Method} & \textbf{\makecell{Basic \\test accuracy}} & \textbf{\makecell{Basic watermark\\ retention rate}} & \textbf{\makecell{Test accuracy}} & \textbf{\makecell{Watermark \\retention rate}} & \textbf{\makecell{Watermark detection\\ capability}} \\
      \hline
      \multirow{3}*{\textbf{MNIST \cite{Deng2012TheMD}}}& REFIT \cite{abs-1911-07205} & 98.08\% & 99.86\% & 97.68\% & 3.27\% & No \\
      \cline{2-7}
      & WILD \cite{Liu2021RemovingBW} & 98.08\% & 99.86\% & 97.57\% & 0.92\% & No \\
      \cline{2-7}
      &Ours & 99.59\%  &  99.93\% & 98.57\% & 1.2\% & Yes \\
      \hline
      \multirow{3}*{\textbf{CIFAR10 \cite{Krizhevsky2009LearningML}}}&REFIT \cite{abs-1911-07205} & 91.31\% & 99.38\% & 89.27\% &  1.71\% & No \\
      \cline{2-7}
      & WILD \cite{Liu2021RemovingBW} & 91.31\% &  99.38\% & 86.16\% & 2.78\% & No \\
      \cline{2-7}
      &Ours & 86.11\% & 99.99\% & 83.44\% & 1.4\% & Yes \\
    \hline
    \end{tabular}
    \label{tab6}
\end{center}
\end{table*}

As shown in Table \ref{tab6}, the watermark retention rates on the MNIST \cite{Deng2012TheMD} dataset are 3.27\% (REFIT), 0.92\% (WILD), 1.2\% (Ours), respectively.
The watermark retention rates on the CIFAR10 \cite{Krizhevsky2009LearningML} dataset are 1.71\% (REFIT), 2.78\% (WILD), 1.4\% (Ours), respectively.
It is shown that, our attack has similar watermark removal capabilities as both REFIT and WILD when the same amount of training data is used.
However, our method can achieve a similar watermark removal effect when only 5\% of the training data is used, which is shown in Table \ref{tab4} and Table \ref{tab5}.
The test accuracy of the compared three methods is reduced by 1\%$\sim$5\%.
However, unlike REFIT and WILD, the proposed attack has the unique capability of detecting watermarks. Specifically, in the proposed attack, the attacker can obtain information about the target class of the backdoor-based watermarking method, and the approximate position and shape of the backdoor trigger in images.
The capability of detecting watermarks is equivalently as significant as the capability of removing watermarks.
Using the information obtained by watermark detection, the attacker can not only efficiently remove the watermark in the DNN, but also develop an effective countermeasure against copyright protectors.

\section{Conclusion} \label{conclusion}
Existing DNN watermarking methods are vulnerable to watermark removal attacks.
For the first time, this paper presents a GAN-based watermark removal method.
The proposed attack consists of two stages. In the first stage, an attacker utilizes GAN to detect and reverse potential watermark triggers in the DNN model.
In the second stage, the attacker uses the reverse watermark trigger pattern to retrain the watermarked model.
The proposed attack can effectively remove the DNN watermark.
Experimental results show that, under the proposed attack, the watermark retention rate can be reduced by 98\%.
In the meantime, the test accuracy of the DNN model will not be affected significantly.
An obvious advantage of the proposed attack is that, compared with the existing watermark removal methods \cite{abs-1911-07205, Liu2021RemovingBW}, the attacker can remove the backdoor-based watermark in the DNN with fewer training samples, and can reverse the backdoor trigger.
This work reveals the fragility of the current backdoor-based watermarking methods and provides guidance for the design of effective copyright protection techniques for deep learning models.

\ifCLASSOPTIONcaptionsoff
  \newpage
\fi

\bibliographystyle{IEEEtran}
\bibliography{paper}

\begin{thebibliography}{10}
\providecommand{\url}[1]{#1}
\csname url@samestyle\endcsname
\providecommand{\newblock}{\relax}
\providecommand{\bibinfo}[2]{#2}
\providecommand{\BIBentrySTDinterwordspacing}{\spaceskip=0pt\relax}
\providecommand{\BIBentryALTinterwordstretchfactor}{4}
\providecommand{\BIBentryALTinterwordspacing}{\spaceskip=\fontdimen2\font plus
\BIBentryALTinterwordstretchfactor\fontdimen3\font minus
  \fontdimen4\font\relax}
\providecommand{\BIBforeignlanguage}[2]{{%
\expandafter\ifx\csname l@#1\endcsname\relax
\typeout{** WARNING: IEEEtran.bst: No hyphenation pattern has been}%
\typeout{** loaded for the language `#1'. Using the pattern for}%
\typeout{** the default language instead.}%
\else
\language=\csname l@#1\endcsname
\fi
#2}}
\providecommand{\BIBdecl}{\relax}
\BIBdecl

\bibitem{RibeiroGC15}
M.~Ribeiro, K.~Grolinger, and M.~A.~M. Capretz, ``{MLaaS}: {M}achine learning
  as a service,'' in \emph{14th {IEEE} International Conference on Machine
  Learning and Applications}, 2015, pp. 896--902.

\bibitem{Adi2018TurningYW}
Y.~Adi, C.~Baum, M.~Ciss{\'{e}}, B.~Pinkas, and J.~Keshet, ``Turning your
  weakness into a strength: {W}atermarking deep neural networks by
  backdooring,'' in \emph{27th {USENIX} Security Symposium}, 2018, pp.
  1615--1631.

\bibitem{Zhang2018ProtectingIP}
J.~Zhang, Z.~Gu, J.~Jang, H.~Wu, M.~P. Stoecklin, H.~Huang, and I.~M. Molloy,
  ``Protecting intellectual property of deep neural networks with
  watermarking,'' in \emph{Proceedings of the Asia Conference on Computer and
  Communications Security}, 2018, pp. 159--172.

\bibitem{Merrer2019AdversarialFS}
E.~L. Merrer, P.~P{\'{e}}rez, and G.~Tr{\'{e}}dan, ``Adversarial frontier
  stitching for remote neural network watermarking,'' \emph{Neural Computing
  and Applications}, vol.~32, no.~13, pp. 9233--9244, 2020.

\bibitem{Uchida2017EmbeddingWI}
Y.~Uchida, Y.~Nagai, S.~Sakazawa, and S.~Satoh, ``Embedding watermarks into
  deep neural networks,'' in \emph{Proceedings of the {ACM} on International
  Conference on Multimedia Retrieval}, 2017, pp. 269--277.

\bibitem{abs-1911-07205}
X.~Chen, W.~Wang, C.~Bender, Y.~Ding, R.~Jia, B.~Li, and D.~Song, ``{REFIT:} a
  unified watermark removal framework for deep learning systems with limited
  data,'' \emph{arXiv:1911.07205}, 2019.

\bibitem{Liu2021RemovingBW}
X.~Liu, F.~Li, B.~Wen, and Q.~Li, ``Removing backdoor-based watermarks in
  neural networks with limited data,'' in \emph{25th International Conference
  on Pattern Recognition}, 2020, pp. 10\,149--10\,156.

\bibitem{Shafieinejad2019OnTR}
M.~Shafieinejad, J.~Wang, N.~Lukas, and F.~Kerschbaum, ``On the robustness of
  the backdoor-based watermarking in deep neural networks,''
  \emph{arXiv:1906.07745}, 2019.

\bibitem{Aiken2020NeuralNL}
W.~Aiken, H.~Kim, and S.~S. Woo, ``Neural network laundering: {R}emoving
  black-box backdoor watermarks from deep neural networks,''
  \emph{arXiv:2004.11368}, 2020.

\bibitem{Yang2019EffectivenessOD}
Z.~Yang, H.~Dang, and E.~Chang, ``Effectiveness of distillation attack and
  countermeasure on neural network watermarking,'' \emph{arXiv:1906.06046},
  2019.

\bibitem{Wang2019NeuralCI}
B.~Wang, Y.~Yao, S.~Shan, H.~Li, B.~Viswanath, H.~Zheng, and B.~Y. Zhao,
  ``Neural cleanse: {I}dentifying and mitigating backdoor attacks in neural
  networks,'' in \emph{{IEEE} Symposium on Security and Privacy}, 2019, pp.
  707--723.

\bibitem{Goodfellow2014GenerativeAN}
I.~J. Goodfellow, J.~Pouget{-}Abadie, M.~Mirza, B.~Xu, D.~Warde{-}Farley,
  S.~Ozair, A.~C. Courville, and Y.~Bengio, ``Generative adversarial nets,'' in
  \emph{Advances in Neural Information Processing Systems}, 2014, pp.
  2672--2680.

\bibitem{Deng2012TheMD}
L.~Deng, ``The {MNIST} database of handwritten digit images for machine
  learning research [best of the web],'' \emph{{IEEE} Signal Processing
  Magazine}, vol.~29, no.~6, pp. 141--142, 2012.

\bibitem{Krizhevsky2009LearningML}
A.~Krizhevsky, G.~Hinton \emph{et~al.}, ``Learning multiple layers of features
  from tiny images,'' Technical Report, 2009.

\bibitem{TramerZJRR16}
F.~Tram{\`{e}}r, F.~Zhang, A.~Juels, M.~K. Reiter, and T.~Ristenpart,
  ``Stealing machine learning models via prediction {APIs},'' in \emph{25th
  {USENIX} Security Symposium}, 2016, pp. 601--618.

\bibitem{XiaoLZHLS18}
C.~Xiao, B.~Li, J.~Zhu, W.~He, M.~Liu, and D.~Song, ``Generating adversarial
  examples with adversarial networks,'' in \emph{Proceedings of the 27th
  International Joint Conference on Artificial Intelligence}, 2018, pp.
  3905--3911.

\bibitem{Zhu2020GangSweepSO}
L.~Zhu, R.~Ning, C.~Wang, C.~Xin, and H.~Wu, ``{GangSweep}: {S}weep out neural
  backdoors by {GAN},'' in \emph{The 28th {ACM} International Conference on
  Multimedia}, 2020, pp. 3173--3181.

\bibitem{LeCun1998GradientbasedLA}
Y.~LeCun, L.~Bottou, Y.~Bengio, and P.~Haffner, ``Gradient-based learning
  applied to document recognition,'' \emph{Proceedings of the {IEEE}}, vol.~86,
  no.~11, pp. 2278--2324, 1998.

\bibitem{He2016DeepRL}
K.~He, X.~Zhang, S.~Ren, and J.~Sun, ``Deep residual learning for image
  recognition,'' in \emph{{IEEE} Conference on Computer Vision and Pattern
  Recognition}, 2016, pp. 770--778.

\end{thebibliography}

\end{document}